\documentclass[aps, prd, twocolumn, preprintnumbers,nofootnoteinbib]{revtex4-1}
\usepackage[T1]{fontenc}
\usepackage{microtype}
\usepackage[textsize=scriptsize,backgroundcolor=red!70,linecolor=red]{todonotes}
\usepackage{booktabs}
\usepackage{dcolumn}
\usepackage{amsmath}
\usepackage{amsfonts}
\usepackage{amssymb}
\usepackage{graphicx}
\usepackage{hyperref}
\usepackage[all]{hypcap}
\usepackage{paralist}
\usepackage{multirow}

\usepackage{graphicx}
\usepackage{dcolumn}
\usepackage{bm}
\usepackage{epsf}
\usepackage{dcolumn}
\def\be{\begin{equation}}
\def\ee{\end{equation}}
\def\bea{\begin{eqnarray}}
\def\eea{\end{eqnarray}}\def\nn{\nonumber}
\def\gsim{\ \rlap{\raise 2pt\hbox{$>$}}{\lower 2pt \hbox{$\sim$}}\ }
\def\lsim{\ \rlap{\raise 2pt\hbox{$<$}}{\lower 2pt \hbox{$\sim$}}\ }
\def\dslash{\kern-4pt \not{\hbox{\kern-2pt $\partial$}}}
\def\pslash{\not{\hbox{\kern-2pt p}}}

\def\l{{\rm L}}

\def\bsmumu{b \to s \mu^+ \mu^-}

\def \kstar{K^*}
\def\BKstarmumu{\bd \to \kstar \mu^+ \mu^-}
\def\bra#1{\left\langle #1\right|}
\def\ket#1{\left| #1\right\rangle}

\def \be{\beta}

\def\bdbar{{\bar B}^0}

\def\bsbar{{\bar B}^0_s}

\def\beq{\begin{equation}}
\def\eeq{\end{equation}}
\def\bea{\begin{eqnarray}}
\def\eea{\end{eqnarray}}
\def\ber{\begin{eqnarray*}}
\def\eer{\end{eqnarray*}}
\def\bwt{\begin{widetext}}
\def\ewt{\end{widetext}}
\def\nn{\nonumber}
\def\roughly#1{\mathrel{\raise.3ex\hbox
{$#1$\kern-.75em\lower1ex\hbox{$\sim$}}}}
\def\lsim{\roughly<}
\def\gsim{\roughly>}

\def\bdbar{{\bar B}^0}

\def\order{\lower 1.8ex \hbox{\LARGE\~{}}}

\usepackage{bm}

\def \kstar{K^*}
\def\BKstarmumu{\bd \to \kstar \mu^+ \mu^-}
\def\bra#1{\left\langle #1\right|}
\def\ket#1{\left| #1\right\rangle}

\def \({\left(}
\def \){\right)}
\def \[{\left[}
\def \]{\right]}
\def \l|{\left|}
\def \r|{\right|}

\def \nn{\nonumber}

\def \be{\beta}

\def\Z{Z^{\prime}}

\def\bsmumu{ b \to  s \mu^+ \mu^-}

\def \kstar{{\bar{K}^*}}
\def\Bsmumu{\bsbar \to \mu^+ \mu^-}

\def\BKmumu{\bdbar \to {\bar K} \mu^+ \mu^-}
\def\BKstarmumu{\bdbar\to \kstar \mu^+ \mu^-}
\def\BXsmumu{\bdbar \to X_s \mu^+ \mu^-}

\def\bdbar{{\bar B}_d^0}
\def\bsbar{{\bar B}_s^0}

\begin{document}
\DeclareGraphicsExtensions{.eps,.ps}


\title{A light $Z^\prime$ for the $R_K$ puzzle and nonstandard neutrino interactions}


\author{Alakabha Datta}
\affiliation{Department of Physics and Astronomy, University of Mississippi, 108 Lewis Hall, Oxford, MS 38677, USA}
\affiliation{Department of Physics and Astronomy, University of Hawaii at Manoa, Honolulu, HI 96822, USA}

\author{Jiajun Liao}
\affiliation{Department of Physics and Astronomy, University of Hawaii at Manoa, Honolulu, HI 96822, USA}
 
\author{Danny Marfatia}
\affiliation{Department of Physics and Astronomy, University of Hawaii at Manoa, Honolulu, HI 96822, USA}

\begin{abstract}

We show that the $R_K$ puzzle in LHCb data and the discrepancy in the anomalous magnetic moment of the muon can be simultaneously explained if a 10~MeV mass $Z^\prime$ boson couples to the muon but not the electron, and that clear evidence of the nonstandard matter interactions of neutrinos induced by this coupling may be found at DUNE.
  
\end{abstract}
\maketitle

There are several perplexing anomalies related to the muon including its anomalous magnetic moment~\cite{Bernreuther:1990jx} and the charge radius of the proton extracted from muonic hydrogen~\cite{Pohl:2010zza}.
In $B$ physics, data from $ b \to s \ell \ell$ decays   indicate evidence of lepton flavor universality  -- the so called $R_K$ puzzle.
The LHCb Collaboration has found a hint of lepton
  non-universality in the  ratio $R_K \equiv {\cal B}(B^+
  \to K^+ \mu^+ \mu^-)/{\cal B}(B^+ \to K^+ e^+ e^-)=0.745\pm0.097$ in the dilepton
  invariant mass-squared range 1 GeV$^2$ $\le q^2 \le 6$~GeV$^2$~\cite{RKexpt}. We take 
  the view that the $R_K$ puzzle may also be a consequence of new physics (NP) affecting the muon.
There is also an anomaly in one of the angular observables in $ B \to K^* \mu^+ \mu^-$ decay~\cite{virto} which may be subject to large hadronic uncertainties~\cite{silves}.
However, unlike
the $R_K$ puzzle lepton non-universal new physics is not necessary to explain the anomaly~\cite{Datta:2013kja}. 

Here we focus on the $R_K$ puzzle which is a clean probe of the Standard Model (SM) due to very small hadronic uncertainties. 
Several NP models with heavy mediators have been considered to explain the $R_K$ puzzle.
We consider a simple NP scenario with a $\Z$ lighter than the muon. The $\Z$ has flavor conserving coupling to quarks and leptons and in addition
we assume that there is a flavor-changing $ b s \Z$ vertex. The $\Z$ couplings to the lepton generations are non-universal to solve the $R_K$ puzzle. In particular we assume the $\Z$ has suppressed couplings to first generation leptons
but has non-negligible couplings to second and third generation leptons.
We constrain the $ bs\Z$ coupling using $B \to K \nu \bar \nu$ and $B_s$ mixing and then from $R_K$ we fix the $\Z$ coupling to muons.
We check that the coupling to muons is consistent with the muon $a_\mu \equiv (g-2)_\mu/2$ measurement, $\Delta a_\mu\equiv a_\mu^\text{exp}-a_\mu^\text{SM}=(29\pm 9) \times 10^{-10}$~\cite{Jegerlehner:2009ry}.  $B \to K \nu \bar \nu$   does not fix the $\Z$ coupling to neutrinos, but assuming SU(2) invariance we set the  $\Z$ neutrino couplings to the charged lepton couplings.
Estimates of the  $\Z$ couplings to light quarks are obtained from non-leptonic   
$ b \to s \bar{q} q$ transitions where $q=u,d,s$. After we obtain the constraints from $B$ physics, we study their implications for nonstandard neutrino interactions (NSI) at DUNE~\cite{Acciarri:2015uup}.

{\bf{$\mathbf{bs\Z}$ vertex.}}
We assume there is a light $Z^{\prime}$ with mass of order 10~MeV. 
The most general form of the $ b s \Z$ vertex with vector type coupling is
\bea
H_{bs\Z} & = &  F(q^2) \bar{s} \gamma^{\mu}  b Z^\prime_{\mu}\,,\
\eea
where the form factor $F(q^2)$ can be expanded as
\bea
F(q^2) & = & a_{bs} + g_{bs} \frac{q^2}{m_B^2} + \ldots\,,\
\eea
where $m_B$  is the $B$ meson mass and the momentum transfer $q^2 \ll m_B^2$.
The leading order term $a_{bs}$ is constrained by $B \to K \nu \bar\nu$  to be smaller than $10^{-9}$~\cite{george}. As will become clear below, the solution to the $R_K$ puzzle would then require the $\Z$ coupling to muons to be $O(1)$ or larger which is
in conflict with the $(g-2)_\mu$ measurement.
The absence of flavor-changing neutral currents forces $a_{bs} \sim 0$, so that
\bea
H_{bs\Z} & = & g_{bs} \frac{q^2}{m_B^2} \bar{s} \gamma^{\mu} b Z^\prime_{\mu}\,,\
\label{Heff_FC}
\eea
where  $g_{bs}$ is assumed to be real.

{\bf $\mathbf{B \to K\nu\bar{\nu}}$.} 
Assuming Gaussian errors, the 95\% C.L. upper limit for  $ B \to K\nu\bar{\nu}$ is~\cite{BKnunubarBaBar} 
\bea
\mathcal{B}(B \to K\nu\bar{\nu}) & \le & 1.9 \times 10^{-5}\,. \
\label{Knunu}
\eea
{}From Ref.~\cite{Buras:2014fpa},  the SM
prediction is
\ber
 \mathcal{B}(B \to K \nu\bar\nu)_{\rm SM} = (3.98 \pm 0.43 \pm 0.19) \times 10^{-6} \,. \
\eer
%
%

The SM Hamiltonian for each neutrino generation is
\begin{equation}
   H_{eff} = -{4G_F \over \sqrt{2}} V_{tb} V_{ts}^* 
    {\alpha \over 4 \pi\sin^2\theta_W}   
   \left[ C_9^{\nu} O_9^{\nu } +C_{10}^\nu O_{10}^{\nu } \right] \,,
\end{equation}
where
\bea
O_9^\nu & = &  (\bar{s} \gamma^\mu P_L b)(\bar{\nu} \gamma_\mu \nu)\,, \nonumber\\
O_{10}^\nu  & = &  (\bar{s} \gamma^\mu P_L b)    (\bar{\nu} \gamma_\mu \gamma_5 \nu)\,. \ 
\eea
In the SM, the Wilson coefficient is determined by box and $Z$-penguin loop diagrams computation which gives,
\begin{equation}
 C_9^\nu = - C_{10}^\nu =-X(m_t^2/m_W^2) \,, 
\end{equation}
where the loop function $X$ can be found e.g. in Ref.~\cite{Buras:1998raa}.

Now we introduce a $\Z$ coupling only to left-handed neutrinos. We further simplify by assuming only flavor conserving couplings but do not assume the couplings to be generation-independent.
We write for generation $\alpha=\mu,\tau$,
\bea
H_{\nu_\alpha \nu_\alpha \Z} & = & g_{\nu_\alpha \nu_\alpha } \bar{\nu}_{\alpha L} \gamma^{\mu} \nu_{\alpha L} Z^\prime_{\mu}\,,\
\label{Heff_nunu}
\eea

Equations~(\ref{Heff_FC}) and~(\ref{Heff_nunu}) lead to the Hamiltonian for $ b \to s \nu_\alpha \bar{\nu}_\alpha$ decays,
\bea
H_{bs \nu_\alpha \nu_\alpha} & = & 
- \frac{g_{bs} g_{\nu_\alpha \nu_\alpha}^*}
{q^2- m_{\Z}^2}   \frac{q^2}{m_B^2} \bar{s} \gamma^{\mu} b  \bar{\nu}_{\alpha L} \gamma_{\mu} \nu_{\alpha L}\,.\
\eea
%
We get $\mathcal{B}(B \to K\nu\bar{\nu})=3.96\times 10^{-6}$ for the SM.
From Eq.~(\ref{Knunu}) we obtain the $2\sigma$ constraint,
\bea
|g_{bs}| \lsim 1.4\times 10^{-5}\,.
\label{eq:BKnunu}
\eea
Note that this constraint does not dependent on $g_{\nu\nu}$ as the NP contribution is dominated by the two body $ b \to s \Z $ transition.
In principle, we can also consider $ B \to K^*\nu\bar{\nu}$ but only certain helicity amplitudes are affected by NP. Furthermore at low $q^2$ the NP amplitudes are suppressed. Hence this decay provides a weaker constraint than $B \to K\nu\bar{\nu}$.

{\bf  $\mathbf{B_s}$ mixing.}
Absent knowledge of $F(q^2)$ for $q^2 \sim m_B^2$, we assume that effects of the longitudinal polarization of the $Z^\prime$ are compensated by the form factor so that the Hamiltonian responsible for $B_s$ mixing can be written as
\bea
H_{B_s} \approx - \frac{g_{bs}^2}{m_{B_s} ^2- m_{\Z}^2}  \bar{s} \gamma^{\mu} b  \bar{s} \gamma_{\mu} b\,.\
\label{Bs}
\eea

The correction to $B_s$ mixing is given by
\bea
\Delta M_s^{NP} & = &   - \frac{g_{bs}^2}{m_{B_s} ^2- m_{\Z}^2}  \bra{ B_s^0} \bar{s} \gamma^{\mu} b  \bar{s} \gamma_{\mu} b\ket{\bar{B}_s^0}.\
\eea
Using the vacuum insertion approximation~\cite{soni_2HDM} and the fact that $m_{B_s}  \approx m_b+m_s$,
 \bea
\Delta M_s^{NP} &  \approx &  \frac{g_{bs}^2}{m_{B_s} ^2- m_{\Z}^2} 
\frac{1}{3} m_{B_s}  f_{B_s}^2\,. \
\label{NPBs}
\eea


%
%
The mass difference  in the SM is  given by
\beq
 \Delta M_s^{SM} = \frac{2}{3} m_{B_s} f_{B_s}^2 \hat B_{B_s} |N C_{VLL} |,\
\eeq
%
%
%
where
\ber
N & = & \frac{G_F^2 m_W^2}{16\pi^2} (V_{tb} V_{ts}^*)^2 , 
 \nn\\
C_{VLL} & = & \eta_{B_s} x_t \left[ 1 + \frac{9}{1-x_t} - \frac{6}{(1-x_t)^2} -\frac{6 x_t^2 \ln x_t}{(1-x_t)^3} \right] .
\eer
In the above, $x_t \equiv m_t^2/m_W^2$, $\eta_{B_s} = 0.551$ is the
QCD correction \cite{Buchalla:1995vs} and $\hat B_{B_s} $ is the bag parameter. 
%
%
%
Taking $f_{B_s}\sqrt{\hat B_{B_s}} = (266 \pm 18)$~MeV~\cite{Aoki:2013ldr},
$V_{tb} V_{ts}^* = -0.0405 \pm 0.0012$~\cite{pdg,Charles:2015gya}, and $\overline{m}_t
= 160$~GeV~\cite{pdg,Chetyrkin:2000yt},  the SM prediction is \cite{newpaper_datta}
\beq
\Delta M_s^{SM} = (17.4 \pm 2.6)~{\rm ps}^{-1} ~.
\eeq
This is to be compared with the experimental measurement~\cite{HFAG},
\beq
\Delta M_s = (17.757 \pm 0.021)~{\rm ps}^{-1}\,,
\eeq
which is consistent with the SM prediction. To bound the NP coupling $g_{bs}$ we take the NP contribution to be at most the 1$\sigma$ uncertainty in the SM contribution, i.e., 
$\Delta M_s^{NP} \sim 2. 6~$ps$^{-1}$.  
With the $B_s$ decay constant $f_{B_s}$ from Ref.~\cite{fBs}, and assuming $m_{B_s} \gg m_{\Z}$, Eq.~(\ref{NPBs}) yields
\bea
|g_{bs}| & \lsim & 2.3\times 10^{-5} .\ 
\label{eq:gbs}
\eea
This is consistent with the bound obtained on $g_{bs}$ from $ B \to K \nu \bar{\nu}$.

{\bf $\mathbf{R_K}$ puzzle.}
Here we follow the discussions in Ref.~\cite{jhepcpc, jhepcpv}. Within the SM, the effective Hamiltonian for the quark-level
transition $\bsmumu$ is  \cite{Buras:1994dj}
\bea
{\cal H}_{\rm eff}^{SM} &=& -\frac{4 G_F}{\sqrt{2}}
\, V_{ts}^* V_{tb} \, 
\Bigl\{ \sum_{i=1}^{6} {C}_i (\mu) {\cal O}_i (\mu) \nn \\
&&+ C_7 \,\frac{e}{16 \pi^2}\, [\bar{s}
  \sigma_{\mu\nu} (m_s P_L + m_b P_R) b] \,
F^{\mu \nu} \nn \\
&& +\, C_9 \,\frac{\alpha_{em}}{4 \pi}\, (\bar{s}
\gamma^\mu P_L b) \, \bar{\mu} \gamma_\mu \mu  \nn \\
&&+ C_{10}\,\frac{\alpha_{em}}{4 \pi}\, (\bar{s} \gamma^\mu P_L b) \, \bar{\mu}
\gamma_\mu \gamma_5
\mu  \, \Bigr\} ~,
\label{HSM}
\eea
where $P_{L,R} = (1 \mp \gamma_5)/2$. The operators ${\cal O}_i$
($i=1\ldots 6$) correspond to the $P_i$ in Ref.~\cite{bmu}, and $m_b =
m_b(\mu)$ is the running $b$-quark mass in the $\overline{\rm MS}$
scheme.  We use the SM Wilson coefficients as given in
Ref.~\cite{Altmannshofer:2008dz}.  

\begin{figure}
	\centering
	\includegraphics[width=0.45\textwidth]{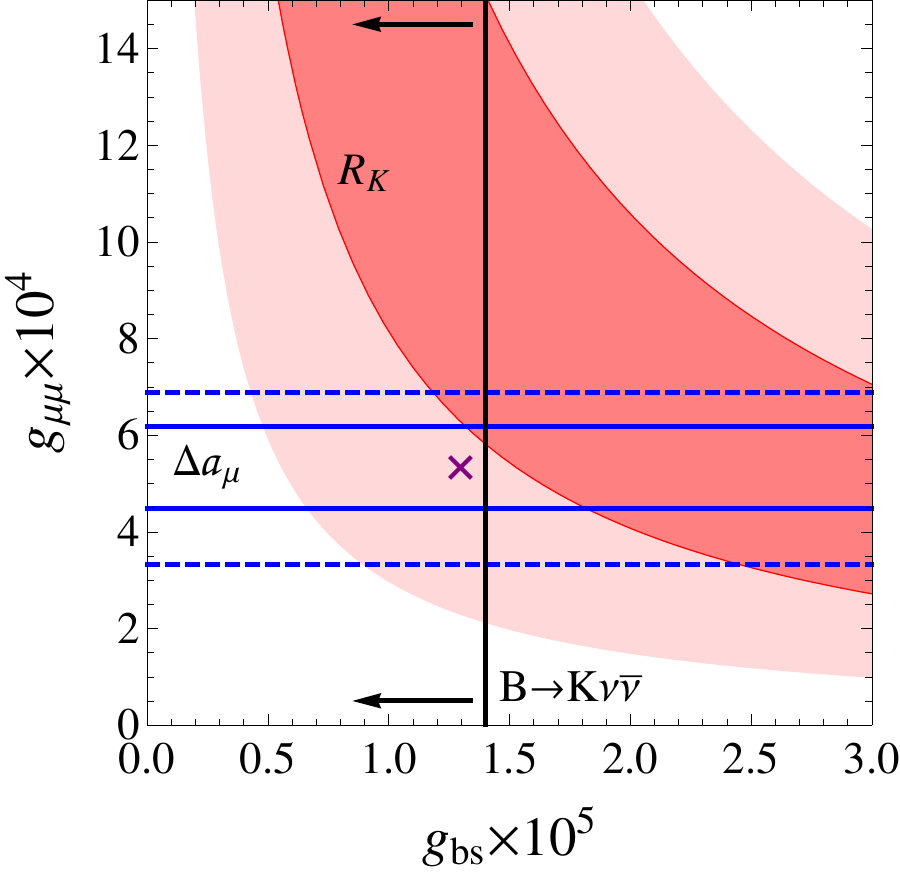}
	\caption{The allowed regions in the ($g_{bs}$, $g_{\mu\mu}$) plane for $m_{\Z}=10$~MeV. The shaded bands are the 1$\sigma$ and 2$\sigma$ regions favored by  $R_K$. The regions between the horizontal solid and dashed lines explain the discrepancy in the anomalous magnetic moment of the muon at the 
	1$\sigma$ and 2$\sigma$ C.L. The vertical line shows the 2$\sigma$ upper limit on $g_{bs}$ from $B \to K\nu\bar{\nu}$. The cross denotes the parameters used for studying neutrino NSI.}
	\label{fig:couplings}
\end{figure}

Introducing a $\Z$ coupling to leptons
\bea
H_{\ell \ell \Z} & = & g_{\ell \ell } \bar{\ell} \gamma^{\mu} \ell Z^\prime_{\mu},\
\label{Heff_ll}
\eea
Equations~(\ref{Heff_FC}) and~(\ref{Heff_ll}) lead to the Hamiltonian for $ b \to s \ell \ell$ decays
\bea
H_{bs \ell \ell} & = & 
- \frac{g_{bs} g_{\ell \ell}^*}
{q^2- m_{\Z}^2}   \frac{q^2}{m_B^2} \bar{s} \gamma^{\mu} b  \bar{\ell} \gamma_{\mu} \ell\,.\
\label{eq:Hbsll}
\eea
We can rewrite this as,
\bea
H_{bs \ell \ell} & = & 
-\frac{4 G_F}{\sqrt{2}}
\, V_{ts}^* V_{tb} \, 
\frac{\alpha_{em}}{4 \pi}\left[
R_V^\ell(q^2)
   \bar{s} \gamma^{\mu}P_L b  \bar{\ell} \gamma_{\mu} \ell  \right.\nn \\
&& \left.+ R_V^{ \prime \ell}(q^2)
   \bar{s} \gamma^{\mu}P_R b  \bar{\ell} \gamma_{\mu} \ell \right]\,,\
\label{bsllfinal}
\eea
where
\bea
R_V^\ell(q^2) & = &   R_V^{\prime \ell}(q^2)=
 \frac{\sqrt{2} \pi g_{bs} g_{\ell \ell}^*}
 {
  G_F V_{ts}^* V_{tb} 
\alpha_{em}}
  \frac{q^2}{m_B^2}     \frac{1}{q^2- m_{\Z}^2}\,.\
\eea
We assume the $\Z$ does not couple to electrons and so ${\cal B}(B^+ \to K^+ e^+ e^-)$ is described by the SM, while ${\cal B}(B^+ \to K^+ \mu^+ \mu^-)$ is modified by NP. We scan the parameter space of $g_{bs}$ and $g_{\mu\mu}$ for values that are consistent with the experimental measurement of $R_K$; see
Fig.~\ref{fig:couplings}.

%
{\bf Muon magnetic moment.} The light $\Z$ also explains the discrepancy in the muon magnetic moment measurement. From Ref.~\cite{Leveille:1977rc}, we have
\bea
\Delta a_\mu =
{ {(g_{\mu\mu})}^2\over 8\pi^2}
\int_0^1 { 2x^2(1-x) \over x^2+ (m_{\Z}^2/m_\mu^2) (1-x)}dx\,.
\eea
For $m_{\Z}=10$~MeV, the measured value of $\Delta a_\mu$ gives $g_{\mu\mu}$ as in Fig.~\ref{fig:couplings}.

{\bf Other constraints.}
We now check that the result is consistent with other  $\bsmumu$ transitions. 
Note that our light $\Z$ cannot be produced as a resonance in $\bsmumu$ decays. Also, as we have a vector coupling in Eq.~(\ref{bsllfinal}) there is no contribution to $\Bsmumu$.

The BaBar Collaboration measures $\mathcal{B}(\bar{B} \to X_s \mu^+ \mu^-) = (0.66\pm 0.88) \times 10^{-6}$ in the range \mbox{1~GeV$^2$ $\le q^2 \le 6$~GeV$^2$}~\cite{Lees:2013nxa}.
The differential branching ratio for $\BXsmumu$  with SM and the general NP operators can be
found in  Ref.~\cite{jhepcpc}.
We find that the NP contribution to $\mathcal{B}(\BXsmumu$) is only 7\% of the SM prediction for 1~GeV$^2$ $\le q^2 \le 6$~GeV$^2$. Given the current experimental uncertainties, the constraint from this decay is not stringent.
%
%

The branching fractions for $\BKmumu$, $\BKstarmumu$ and the corresponding electron modes
 are known for the entire kinematical range. However due to the long distance contributions we do not use  them to directly constrain NP. 

Finally, the NP amplitude for $\BKstarmumu$ in the low $q^2$ region is suppressed  relative to the leading SM amplitudes
by  ${ \sqrt{q^2} \over m_B}$ and so this decay does not provide any constraints on the NP coupling. 
 We note in passing that constraints from $ b \to s \tau^+ \tau^-$ decays are very 
 weak~\cite{newpaper_datta} and do not produce a meaningful constraint on  the NP coupling $g_{\tau \tau}$.


{\bf  $\mathbf{b \to s  q \bar{q}}$.}
We now consider the $\Z$ coupling to light quarks with a focus on the up and down quarks:
\bea
H_{q q \Z} & = & g_{q q } \bar{q} \gamma^{\mu} q Z^\prime_{\mu}\,.\
\label{Heff_qq}
\eea
It is reasonable for the $\Z$ coupling to quarks to be of the same size as the coupling to the charged leptons, i.e., \mbox{$\sim 10^{-4}$}.
Decays like $B  \to  K \pi $ can constrain the $\Z$ coupling to light quarks. In spite of the hadronic uncertainties approximate bounds are obtainable from these decays.
Equations~(\ref{Heff_FC}) and~(\ref{Heff_qq}) lead to the Hamiltonian for $ b \to s \bar{q} q$ decays, which is similar to Eq.~(\ref{eq:Hbsll}) with $\ell$ replaced by $q$.

The NP  can add to the electroweak contribution in the SM. It is interesting to speculate if such NP can resolve the so called $K-\pi$ puzzle~\cite{dattaKpi}. This is the difference in the direct CP asymmetry in the decays $B^+ \to \pi^0 K^+$ 
  and $B^0 \to \pi^- K^+$.
   It is puzzling that the  leading amplitudes in both decays are the same in the SM while the former decay also gets contributions from a small color and CKM suppressed tree amplitude and the electroweak penguins for the two decays are different. It is possible that new contributions to the electroweak penguins may resolve the puzzle. However, the situation is a bit complicated. First, there are two other relevant decays, 
$B^+ \to \pi^+ K^0$  and  $B^0 \to \pi^0 K^0$,
 and one has to fit to all the decays. Since these are non-leptonic decays one has to account for hadronic uncertainties.
 
 In naive factorization, our NP  does not contribute at leading order to
 $B^+ \to \pi^0 K^+$ as the vector quark current does not produce a pion but can produce a $\rho$ and will thus contribute to   $B^+ \to \rho^0 K^+$ and  $B^0 \to \rho^- K^+$.
 We can always change the chiral structure of the $ \Z$ coupling to quarks to get a leading order contribution  to  $B \to \pi K$. Our intention here is not to resolve the $K -\pi$ puzzle but we
 can estimate the $\Z qq$ coupling in the following way. 
  A reasonable assumption is that NP produces effects of the size of about 10\% of the SM electroweak penguin.

 Both color allowed and color suppressed electroweak penguins are possible in the decay $B^0 \to \rho^0 K^0$, and we can compare these with the NP amplitude.
The ratio of the NP amplitude to the color allowed penguin is
 \bea
 r  & = &\frac{  \bra{ \rho^0 K^0 }H_{NP} \ket{B^0}}
 {  \bra{ \rho^0 K^0} H_{EW}^{SM} \ket{B^0}},\
 \eea
 where   $ H_{EW}^{SM}$ is the color allowed SM electroweak Hamiltonian.  
 Using naive factorization,
 \bea
 r & = &  \frac{  
 g_{bs}(g_{uu}-g_{dd})m_\rho^2
}
 {
  \frac{ G_F}{ \sqrt{2}} a_9 V_{tb}V_{ts}^* \frac{3}{2}  ({m_\rho^2- m_{\Z}^2}) m_B^2}\,,\
 \eea
 where  we have assumed real couplings.
 The factor $a_9= C_9 + \frac{C_{10}}{N_c}$  where $C_{9,10}$ are the Wilson's coefficients and $N_c=3$ is the number of colors. 
 The ratio of the NP amplitude to the color suppressed electroweak penguin is
 \bea
 s  & = &\frac{  \bra{ \rho^0 K^0 }H_{NP} \ket{B^0}}
 {  \bra{ \rho^0K^0} H_{EW}^{SM,C} \ket{B^0}} \nonumber\\
 &=&
  \frac{  
 g_{bs}(g_{uu} -  g_{dd})m_\rho^2
}
 {
  \frac{ G_F}{ \sqrt{2}} a_{10} V_{tb}V_{ts}^*  ({m_\rho^2- m_{\Z}^2}) m_B^2},\
\eea
where  $a_{10}= C_{10} + \frac{C_{9}}{N_c}$   and   $ H_{EW}^{SM,C}$ is the color suppressed SM electroweak Hamiltonian.
Using $a_9( \mu=m_b) \sim -1.22 \alpha_{em}$ and   
  $a_{10}( \mu=m_b) =  0.04 \alpha_{em}$   \cite{neubert-beneke} and
 requiring $|r| \sim 0.1$ we find
 \bea
| g_{bs}(g_{uu} - g_{dd}) |  \sim 1.3 \times 10^{-8}. \
 \eea
 For $|g_{bs}| \sim 10^{-5}$ we get   $|g_{uu} -g_{dd}| \sim 10^{-3}$. As we discuss next, this leads to nonstandard neutrino interactions that are too large.
 On the other hand requiring $|s| \sim 0.1$ gives
 \bea
 |g_{bs} (g_{uu} -g_{dd}) |  \sim 2.8 \times 10^{-10}. \
 \label{eq:gbsguu}
 \eea 
 In this case  $|g_{uu} - g_{dd}| \sim 10^{-5}$.   We will assume that $g_{uu}$ is the same size as $g_{dd}$ and take these couplings to be $\sim 10^{-5}$ to discuss neutrino NSI.

\begin{figure}
	\centering
	\includegraphics[width=0.4\textwidth]{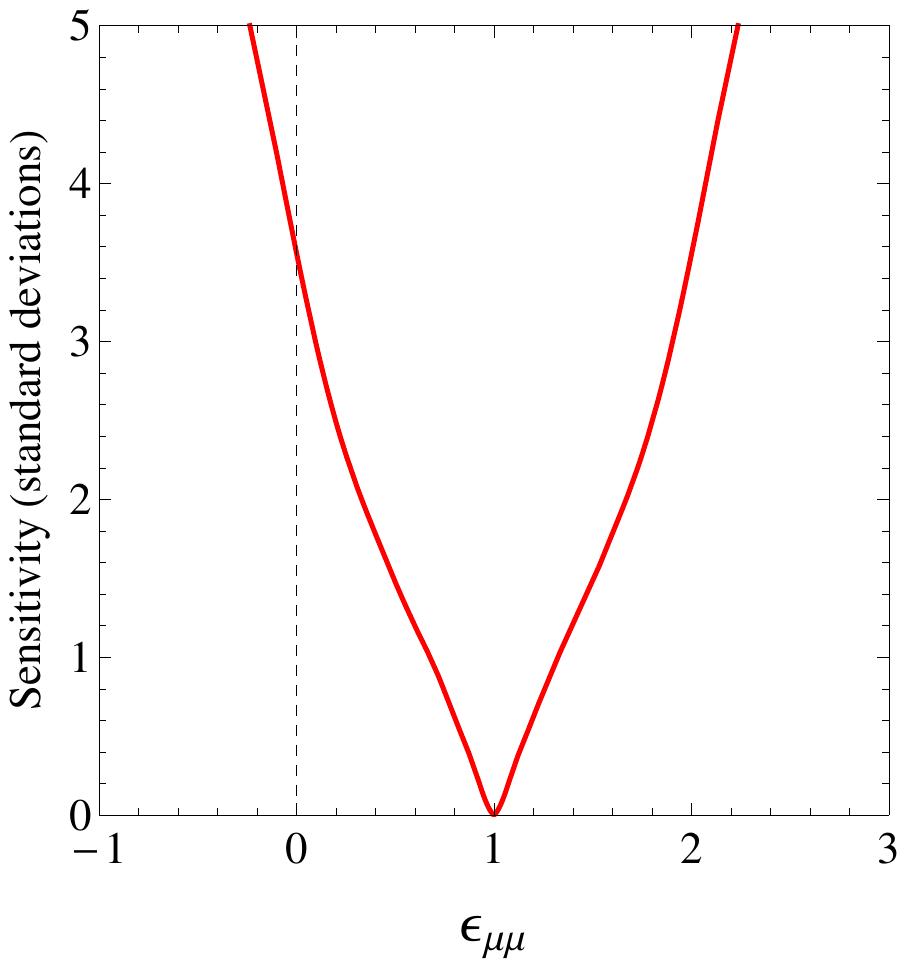}
	\caption{The sensitivity to $\epsilon_{\mu\mu}$ at DUNE. The data are simulated for the normal neutrino mass hierarchy, the neutrino CP phase $\delta=0$, and $\epsilon_{\mu\mu}=1.0$. We assume $\epsilon_{\mu\mu}=\epsilon_{\tau\tau}$.}
	\label{fig:dune}
\end{figure}

{\bf NSI at DUNE.}
The light $\Z$ couplings to neutrinos and first generation quarks affect the neutrino propagation in matter.   The matter NSI can be parameterized by the effective Lagrangian~\cite{Wolfenstein:1977ue},
\bea
  \mathcal{L} =-2\sqrt{2}G_F
   \epsilon^{qC}_{\alpha\alpha} 
        \left[ \overline{\nu}_\alpha \gamma^{\rho} P_L \nu_\alpha \right] 
        \left[ \bar{q} \gamma_{\rho} P_C q \right] + \text{h.c.}\,,
\label{eq:nsi}
\eea
where $\alpha=\mu, \tau$, $C=L,R$, $q=u,d$, and
$\epsilon^{qC}_{\alpha\alpha}$ are dimensionless parameters that represent the
strength of the new interaction in units of $G_F$. 
 Since neutrino propagation in matter is affected by coherent forward scattering, $\epsilon^q_{\alpha\alpha}\equiv \epsilon^{qL}_{\alpha\alpha}+\epsilon^{qR}_{\alpha\alpha}$, can be written as
\bea
\epsilon^q_{\alpha\alpha} =\frac{g_{qq}g_{\nu_\alpha\nu_\alpha}}{2\sqrt{2}G_F m_{\Z}^2 }\,,
\label{eq:epsf}
\eea
regardless of the $\Z$ mass. 
For propagation in the earth, neutrino oscillation experiments are only sensitive to the combination,
\bea
\epsilon_{\alpha\alpha}\approx 3(\epsilon_{\alpha\alpha}^u+\epsilon_{\alpha\alpha}^d)\,.
\label{eq:eps}
\eea

We now use the light $\Z$ couplings obtained from $B$ physics to study signatures at neutrino oscillation experiments.
We assume $g_{\nu_\mu\nu_\mu}=g_{\mu\mu}$, which is motivated by an SU(2) invariant realization of Eq.~(\ref{eq:nsi}).
We fix $g_{\mu\mu} =5.4 \times 10^{-4}$ and $g_{bs}=1.3 \times 10^{-5}$ to explain both the $R_K$ and muon $g-2$ anomalies; this set of couplings is marked by a cross in Fig.~\ref{fig:couplings}.
 To avoid a finetuned cancellation,
we take $g_{uu}=1.2\times 10^{-5}$ and $g_{dd}=-1.0\times 10^{-5}$, which satisfies the relation in Eq.~(\ref{eq:gbsguu}). For $m_{\Z}= 10$~MeV, these couplings satisfy a plethora of constraints~\cite{Farzan:2015doa}. 
From Eqs.~(\ref{eq:epsf}) and~(\ref{eq:eps}), we get $\epsilon_{\mu\mu}=1.0$. To satisfy constraints from current neutrino oscillation data~\cite{Gonzalez-Garcia:2013usa}, we assume $\epsilon_{\tau\tau}=\epsilon_{\mu\mu}$. 

Following the procedure in Ref.~\cite{Liao:2016hsa}, we simulate 300~kt-MW-years of DUNE data with the normal neutrino mass hierarchy, the neutrino CP phase $\delta=0$, and $\epsilon_{\mu\mu}=1.0$. We scan over both the mass hierarchies, the neutrino oscillation parameters and $\epsilon_{\mu\mu}$.  The expected sensitivity of DUNE to reject the SM scenario is shown in Fig.~\ref{fig:dune}.  We see that the SM scenario with $\epsilon_{\mu\mu}=0$ is ruled out at the 3.6$\sigma$ C.L. at DUNE. 

{\bf Summary.} 
We showed that the $R_K$ puzzle in LHCb data can be explained by a light $\Z$ . The resulting coupling of the $\Z$ to muons also reconciles the muon $g-2$ measurement. 
After carefully examining various constraints from $B$ physics, we find that this $\Z$ could yield large NSI in neutrino propagation. We further demonstrated that evidence of NSI induced by the light $\Z$ coupling may be found at DUNE. A scattering experiment at CERN will also search for such a boson~\cite{cern}.

\vskip 0.1in
{\it Acknowledgments.} 
This research was supported by the U.S. NSF under Grant No.
PHY-1414345 and by the U.S. DOE under Grant No. DE-SC0010504.

\vskip1cm


\end{document}